\documentstyle[aps,prl,multicol,epsf]{revtex} 
\begin{document}

\title{Emergent spatial structures in   critical   sandpiles}

\author{Bosiljka Tadi\'c$^{1,*}$ and Deepak Dhar$^{2,**}$}

\address{
$^1$Jo\v{z}ef Stefan Institute, 
P.O. Box 3000, 1001 Ljubljana, Slovenia \\
$^2$ Theoretical Physics Group, Tata Institute of Fundamental Research, 
Homi Bhabha Road, Mumbai 400005, India }

%\date\today

\maketitle
\begin{abstract}
\newline

We introduce and study a  new directed sandpile model with threshold dynamics
and stochastic toppling rules.  We show that particle conservation 
law and the directed percolation-like local evolution of avalanches 
lead to the formation of  a spatial structure in the steady state, 
with the density developing a power law tail away from the top.  
We determine   the scaling exponents characterizing
the avalanche distributions  in terms of  the critical exponents 
of  directed percolation in all dimensions.

\end{abstract}

%\narrowtext
\begin{multicols}{2}

Many extended slowly driven dissipative systems in nature evolve into
self-organized critical (SOC) steady states which show long-range spatial
and temporal correlations. Since the pioneering work of Bak {\it et al.}
\cite{BTW}, sandpile models have served as paradigms of SOC systems.  A
great deal of understanding of SOC has been achieved by numerically
studying sandpile models with different evolution rules \cite{KNWZ}. For a
special subclass of these models, i.e.,  the Abelian Sandpile Models (ASM),
several analytical results are available \cite{ASM}. 
Recent studies of  models with stochastic toppling
rules have shown that these models usually belong to a universality class 
different from  deterministic automata; they  have robust  critical states 
with respect to changes of a control parameter, and  may also exhibit a 
dynamical phase transition between qualitatively different steady 
states \cite{stochastic-spa}.
However, the spatial structures in the steady states of these models are  
much less investigated \cite{structures}.
Due to long-range correlations in the critical states, 
influence of the boundary can be felt deep inside, and this can give 
rise to large-scale spatial structures. 
Indeed, emergent spatial structures are {\it sine qua non}  for a SOC 
theory  of fractals occurring in nature, e.g., mountain landscapes,
river networks, and earthquake fault zones.

In this Letter, we propose a new stochastic sandpile model which
shows  emergent spatial structures in the steady states.
The model is a stochastic generalization of the directed ASM \cite{DR} 
and contains a probabilistic control parameter $p$. 
In the case $p=$1  the exponents are known
exactly in all dimensions \cite{DR}.  We show that for $p \neq 1$, the
model is in a new universality class and 
can be  related to directed percolation (DP) \cite{DP} problem with a
nonuniform concentration profile.  A steady state exists only
for $p > p^\star$, where $p^\star$ equals the critical threshold for
directed site percolation. Above $p^\star $ the system evolves towards
a steady state which is arbitrarily close to  the generalized
DP critical line.
We also show that in the critical state of our model a spatial structure 
results from an interplay of  DP-like local evolution rules on the one 
side, and the dynamic conservation law on the other. We find a power-law 
density profile, which further enables us to determine the exact
 expressions for   the scaling exponents of avalanches in terms of the 
 DP critical exponents in all dimensions $d$. Our numerical simulations 
in two dimensions support these conclusions.

 For concreteness, we consider a square lattice of linear size $L$ 
 with sites $(i,j)$
oriented so that the diagonal direction $(1,1)$ is vertically down.  
A non-negative integer variable, 
height $h(i,j)$, is attached to each lattice site.  Sand grains are added
one at a time at a randomly chosen site on the top layer, 
increasing  its  height by one. 
A site becomes {\it unstable} if $h(i,j) \geq 2$ and 
relaxes as follows:  With probability $p$, the local height
decreases by two, and heights at each of its two downward
neighbors  increase by one. Otherwise, the  heights remain unchanged. 
In either case, the site is considered as stable at the next time step.
 Only sites to which at least one particle was added at the preceding 
 time step are checked for toppling.
A discrete-time parallel update is applied to all such sites.
We apply  periodic boundary conditions in the horizontal direction. 
Two particles leave the system for each toppling
occurring at the bottom layer. 

In the limit $p=1$, the structure of the steady state is known exactly 
\cite{DR}:
Only configurations with heights $h=0$ and $h=1$ occur, and all
such configurations are equally probable.  The avalanche clusters
are compact.
For stochastic toppling, $p \ne 1$, the model has quite a different behavior. 
It is {\it no longer Abelian}, since adding two particles together to a site 
of height $ h\geq 2$ does not have the same effect as adding them one by 
one at different time steps. (In the latter case, toppling can occur twice 
with a finite probability.) With a small probability  heights could
become  arbitrarily large.  The avalanche clusters have branches and 
holes of various sizes. An example  is shown in Fig.\ 1. 

For small $p$ the average
influx of particles per attempted toppling at a site, which is $ \geq 1$,
exceeds the average outflux, $2p$,  thus  there exists a
value $p^\star$, such that a steady state is possible only for $p \geq p^\star$. 
We now argue that $p^\star = p_c$, where $p_c$ is the critical threshold
for the directed site percolation problem \cite{sim}.

Suppose that $\cal{C}$ is a stable configuration of the pile, and a
particle added at the site $(i,j)$ causes on the average $n_\ell(\cal{C})$
topplings at depth $\ell $ below it. If for any stable configuration
$\cal{C' }$, all heights are greater than or equal to corresponding
heights in $\cal{C}$ then $ n_\ell(\cal{C}') \geq$ $n_\ell(\cal{C})$.
Now,  note that for all configurations $\cal{C}$ for which all  sites
have $h\ge 1$, the distribution of size of avalanches is exactly
the same  as in the directed {\it site-}percolation problem on
this lattice.  Therefore, for all $p < p_c$, $n_\ell $ decreases
exponentially with $\ell $. Hence no topplings occur  at large depths, 
and  particles pile up  in the upper layers, and thus there is 
no steady state.  Conversely, for $ p > p_c$, the directed percolation
avalanche clusters typically form a wedge, and $n_\ell$ increases 
with $\ell $ for large $\ell $ (see below). 
Then avalanches in configurations with {\it all} heights $h\geq 1$
cause many topplings, and each layer after the
avalanche on the average loses particles.  Thus if the system
ever reaches a state with large density, the number
of particles in the system will decrease until at sufficiently many sites
heights become low enough so that the propagation of avalanches is affected,
and it becomes {\it critical}, but not {\it super-critical}. Hence the
system will have a steady state  for all $ p \geq p^{\star}$ = $p_c$.
On the square lattice numerical estimate for $p_c$ \cite{DP}
gives $p^\star \approx $ 0.7054853(5).

Our numerical simulations support this conclusion.  
In Fig.\ 2  the probability $P(T)$ that an avalanche has durations 
$\ge T$ is plotted against $T$ for different values of $p$ and $L=$200. 
(Notice that in directed models $T \equiv \ell $). 
For large $T$ it varies as  $P(T)\sim T^{1-\tau _t}$. 
Exponential decay of the lower three curves indicates loss of SOC.
In the steady states for $p\ge p^\star $   the integrated 
cluster-size distribution
behaves as $D(s) \sim s^{1-\tau _s}$, with the following scaling properties 
\begin{equation} 
D(s,L) = L^{1-\tau _t }{\cal{D}}(sL^{-D_\parallel }) \; , 
\label{fss}
\end{equation}
and  the   scaling relation $(\tau _s-1)\ D_\| = 
\tau _t-1$ is satisfied. 
In Fig.\ 3 the distribution $D(s,L)$ vs. $s$ and its 
finite-size-scaling plot are shown for $p=p^\star$.

We now discuss the structure of the steady state for $p^\star \leq p 
\mathopen<1$.
Let $\rho$ be the probability that a site chosen at random  has a nonzero 
height in the steady state. Then the 
probability that this site will topple if a single particle is added to 
it is $P_1 = p\rho$, and the probability that it topples if two 
particles are added to it is $P_2 = p$. The correlations of heights on the same layer are irrelevant 
and can be ignored \cite{correlations}. Therefore, the evolution of an
avalanche  is the same as in a Domany-Kinzel 
(DK) cellular automaton model of directed site-bond percolation \cite{DK}.
 This implies that, in order for the system to have {\it critical} correlations 
in the bulk, the set of points  $(P_1,P_2)$ must lie on the critical 
line of the DK model. 

 However, as we show below, the dynamic conservation law prevents the 
 avalanche clusters  {\it in the steady state}  from being in the universality 
 class of DP. Particle conservation implies that in the steady state
the average number of topplings  at each layer equals 1/2.
On the other hand, in the case of  DP   
 the expected  number of growth sites at depth $\ell$ is known to vary as 
$m \sim \ell^\kappa $ with $\kappa ={{[(d-1)\nu_{\bot}-2\beta}]/{\nu_{\|}}}$,
where $\beta $, $\nu _\|$ and $\nu _\bot $ are standard DP critical exponents
of the order parameter, and parallel and transverse correlation lengths, 
respectively. For DP in $d\mathopen<$5 dimensions   $\kappa \mathclose> 0$, 
thus  $m$ increases with  $\ell $,  which is clearly not possible in the 
steady state.
The way these conflicting requirements of particle conservation and 
locally critical DP-like evolution are satisfied  in 
our model is that the critical steady state develops a spatial structure.
The density $\rho$, and hence $P_1$, are not uniform throughout the 
system, but vary from layer to layer \cite{profiles}. 
Let $\rho(\ell)$ be the average density of sites with non-zero height in the 
$\ell $-th layer. By equating average influx and outflux of particles 
at a site on $\ell $-th
layer, we find that  $ \rho(\ell) = [1-(2p-1)f(\ell)]/[2p (1-f(\ell)]$,
where $f(\ell )$ is the number of topplings caused by simultaneous addition of 
two particles at the site. The exactly  calculated values of $f(\ell )$ 
 for the first few layers indicate that  $\rho (\ell )$ increases 
with $\ell $.  As discussed above, 
for large $\ell$ the profile reaches the value $\rho^{\star}(p) = 
P_1^{\star}(p)/p$, where $(P_1^{\star}(p),p)$ is a point on the DP critical
line in the $(P_1,P_2)$ parameter space. 
In Fig. 4, we  plot the profile $ \rho$ against $\ell$ obtained by  
numerical simulations for $p=p^\star $ and $L=200$. 
The profile  is well described by  a power law: 
\begin{equation}
   \rho(\ell)= \rho^{\star}(p) - A(p) \ell^{-x}\ , 
   \label{rho_ell}
\end{equation}
with $\rho^{\star }=$1 and $A$=0.39 for $p=p^\star$, and $x=$0.578.

We now show  that  the profile given by Eq.\ (\ref{rho_ell}) 
changes the avalanche  statistics in our model, and thus 
strongly affects the bulk transport. In the  bulk, transport of
particles is locally described by the DK model with parameters 
$(P_1(\ell ),P_2)$. For large $\ell$  
 the system is close to the critical line and the local longitudinal 
 correlation length $\xi (\ell)$ varies as
 $\xi (\ell) \sim \left[P_1^\star -P_1(\ell )\right]^{-\nu _\|}$. 
 In order for the transport to propagate  further  to a distance  $\ell $, 
   $\xi (\ell)$  must be proportional to  $\ell $, i.e.,
 \begin{equation}   
\xi (\ell) \approx \ell /B \ . 
\label{xi_ell}
\end{equation}
 This implies that the exponent $x$ in Eq.\ (\ref{rho_ell}) is 
 exactly $x=1/\nu _\|$. 
From the simulation data  in   $\log(1-\rho (\ell ))$ vs. $\log \ell $ 
plot we find the   slope  $x =0.575 \pm 0.005$ (see inset to Fig.\ 4), 
leading to  $\nu _\|   = 1.738 \pm 0.005$, in a good agreement with 
$\nu _\| $ for DP in two-dimensions  \cite{DP}.

The calculation of avalanche exponents for our model reduces to the 
problem of determining the distribution of cluster-sizes of surface 
clusters in a directed  site-bond percolation model where the 
concentration of bonds has a power-law profile (\ref{rho_ell}). 
In the   renormalization-group approach \cite{Binder}  
 $x= 1/\nu_{\|}$ is a marginal case and the  cluster exponents
 may depend on the amplitude  $A$. 

Let $G(R_{\bot},R_{\|})$ be the probability that the site
$(R_{\bot},R_{\|})$ topples
if a particle is added at (0,0) in the steady state.
Since $\rho^{\star}(p)-\rho(\ell ,p) \ll 1$ for large $\ell$, 
we can show that  to leading order of perturbation
\begin{equation}
 G(R_{\bot},R_{\|})\approx G_0(R_{\bot},R_{\|})\ \exp\left[-\int_1^{R_\|} 
 d\ell/\xi(\ell)\right] \ ,
 \label{grr0}
\end{equation}
where $G_0(R_{\bot},R_{\|})$ is the two-point correlation function for the 
critical DP process. Using Eq.\ (\ref{xi_ell}) we get 
\begin{equation}
 G(R_\bot,R_\|)=G_0(R_\bot,R_\|) R_\|^{-B} \ . 
 \label{grrr}
\end{equation} 
The value of B selected by the steady state is determined by the
requirement that the average outflux of particles per avalanche from the
$R_{\|}$-th layer equals one, i.e., $\sum_{R_{\bot}} G(R_{\bot},R_\|) \sim  1$. 
 Since in DP the average outflux is 
$  \sum_{R_{\bot}} G_0(R_{\bot},R_\|) \sim
R_{\|}^{{[(d-1)\nu_{\bot}-2\beta}]/{\nu_{\|}}}$, 
it follows that 
\begin{equation}
     B = {{[(d-1)\nu_{\bot}-2\beta}]/{\nu_{\|}}} \ , 
     \label{B_exps}
\end{equation}
where $\beta $, $\nu _\|$ and $\nu _\bot $ are as above the DP exponents.
Thus  both the power-law tail and the amplitude  $B$ are expressed in 
terms of standard DP exponents.
These facts, in turn, determine the statistics  of
avalanche clusters. In addition to  
 the exponents  for the  distributions of 
duration, $\tau_t$, and size, $\tau_s$,  we also define the 
 anisotropy exponent $\zeta $ for the  average  transverse extent 
 $R_\bot \sim \ell ^\zeta $ of a cluster of length   $\ell$.
 Near the DP critical line $R_\bot $  is expected to have  the  
 scaling behavior as   
\begin{equation}
R_\bot \sim \ell^{\zeta _{DP}} 
\phi ([P_1^\star -P_1(\ell)]\ell ^{1/\nu_\|}) \ .   
\label{trlength}
\end{equation}  
 Notice that due to the power-law profile (\ref{rho_ell}),
the argument of the scaling function $\phi $ in Eq.\ (\ref{trlength})
remains finite in the limit $\ell \to \infty $, and thus  
 $\zeta = \zeta _{DP} = \nu _\bot /\nu _\|$.

In the critical DP, the probability that a perturbation survives up to 
layer $T$ varies as $P_0(T) \sim T^{-\beta/\nu_{\|}}$. As an expression 
similar to Eq.\ (\ref{grr0}) applies also to the survival probability 
in the steady state of our model, we have that  $P(T)=P_0(T)T^{-B}$.    
Using $B$ from Eq.\ (\ref{B_exps}) this gives 
\begin{equation}
  \tau_t = 1+ (d-1)\zeta - \beta/\nu_{\|} \ . 
  \label{tau_t}
\end{equation}
Using standard scaling arguments for the directed SOC system \cite{DR}
we notice first that the expected number of topplings in a cluster
of length $\ell $ scales as $\mathopen<s\mathclose>_\ell \sim 
\ell^{\tau_t}$, that is, $D_\|=\tau_t$. Together with $(\tau_s-1)D_\|
=\tau_t-1$ this then  gives the renormalized $\tau _s$ exponent  as
\begin{equation}
  \tau_s = 2-1/\tau_t \ .
 \label{taus}
\end{equation}
\noindent
Inserting  the best known numerical values of the
exponents  for two-dimensional DP \cite{DP}, 
gives  $ \tau_t $= 1.47244,   $\tau_s $
= 1.32059, and $\zeta $=0.63261 . 
We have checked these predictions against  numerical 
simulations of the exponents and fractal dimension $D_\|$. 
In the inset to Fig.\ 2  various scaling exponents are plotted
versus $p$ in the scaling region $p\ge p^\star $. 
Away from a small crossover region near the point $p=1$, 
 the obtained values of the exponents are independent of $p$ within 
 numerical error. We find 
$\tau_t =$1.460 $\pm $0.014,  $\tau_s $= 1.313 $\pm $ 0.012, and 
$\zeta $= 0.624 $\pm $0.014 in fair agreement with the above conclusions. 

For $d=$3  using Eqs.\ (\ref{tau_t}-\ref{taus}) and known numerical  
values of DP exponents \cite{Grassberger} we find  $\tau _t=$1.674 
and $\tau _s=$1.403. 
The upper critical dimension of our stochastic model is $d_c=$5, in 
contrast to $d_c=$3 in the deterministic limit $p=$1. 
For $d\geq $5, the DP critical exponents are $\beta =$1, $\nu_\|=$1,
and $\nu_\bot =$1/2, leading to  $B=$0, and thus the  exponents have  the
mean-field values $\tau _t=$2, and $\tau _s$=3/2.

In conclusion, we have demonstrated that nearness of the steady  states to 
 the directed-percolation critical line and  the conservation of number 
 of particles in the bulk are responsible 
for the emergent spatial structures in our   stochastic sandpile model.
A power-law density profile has been found and the self-organized criticality
which  is in a different universality class from the deterministic limit.
In all dimensions $d$ the scaling exponents of avalanches have been determined 
in terms of standard directed percolation critical exponents.

The work of B.T. was  supported by the Ministry 
of Science and Technology of the Republic of Slovenia. B.T.  
thanks Al Corral for his assistance in the graphic program.
D.D. would like to acknowledge useful comments on the manuscript
by  M. Barma and S.N. Majumdar.

\narrowtext
\begin{figure}[thb]
\epsfxsize=82mm\epsffile[18 236 592 556]{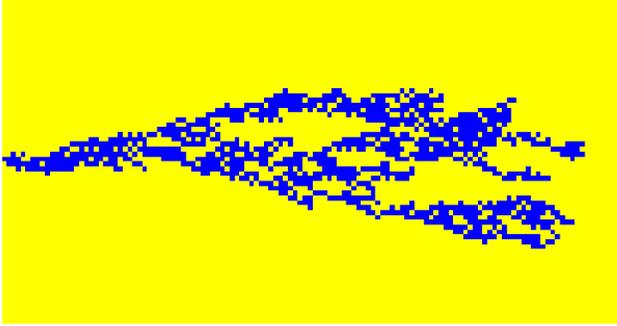}
\caption{\label{fig1} Example of an avalanche running from left to right
(dark) for $p=p^\star $ on the lattice of linear size $L$=128 .}
\end{figure}

\begin{figure}
\epsfxsize=82mm\epsffile[34 68 536 600]{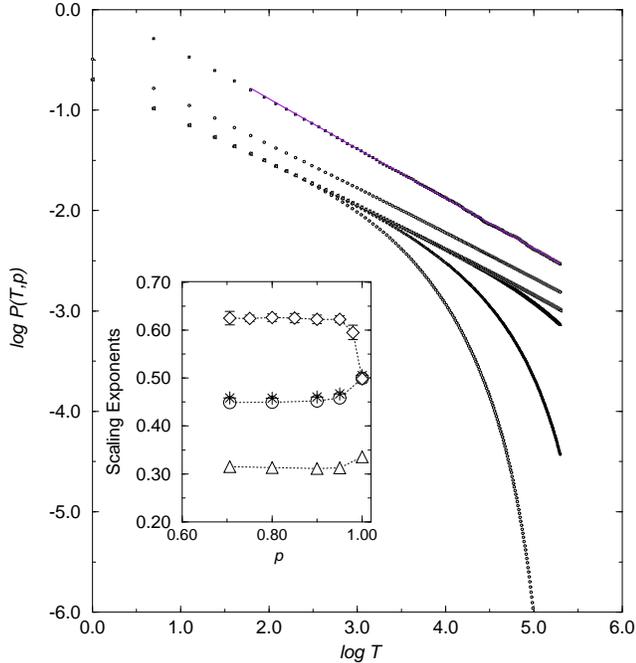}
\caption{\label{fig2}Double logarithmic plot of the integrated
distribution of durations $P(T,p)$ vs.  $T$ for $L$=200, 
and $p$=1, 0.8, 0.70548, 0.69, 0.68, and 0.65 (top to bottom). 
First two curves have been shifted vertically for easier display. 
Inset: Scaling exponents: ($\diamond $) $\zeta $, ($\bigcirc $) 
$\tau_t-1$,
($\triangle $) $\tau_s-1$, and ($\star $) $D_\|(\tau_s-1)$ plotted against 
$p$ in the scaling region.}
\end{figure}

\begin{figure}
\epsfxsize=82mm\epsffile[44 68 536 600]{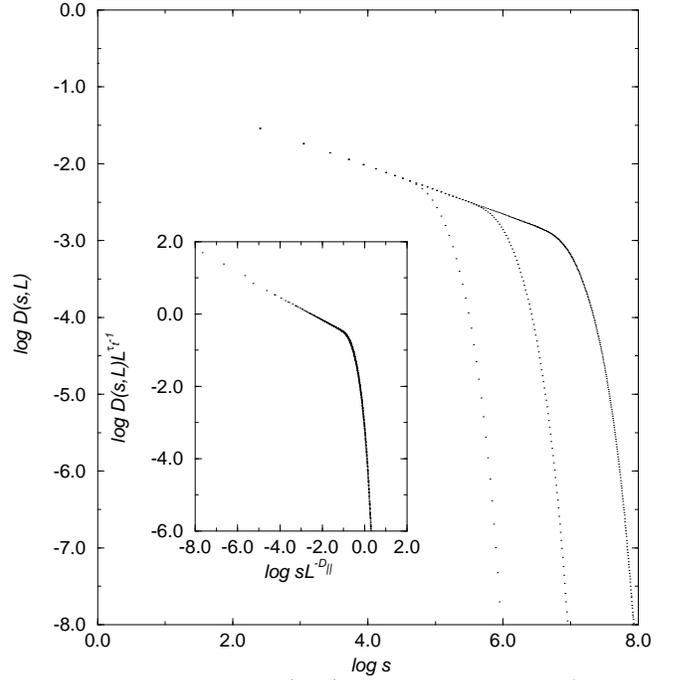}
\caption{\label{fig3}Plot of  $\log D(s,L)$ vs.  $\log s$ for 
$p=p^\star $ and for
three different lattice sizes $L$=50, 100, 
and 200 .  Inset: Data collapse according to Eq.\ (\ref{fss}), where
we used the  values  $\tau_t-1 $=0.45 and $D_\| $=1.45 .}
\end{figure}

\begin{figure}
\epsfxsize=82mm\epsffile[40 68 536 600]{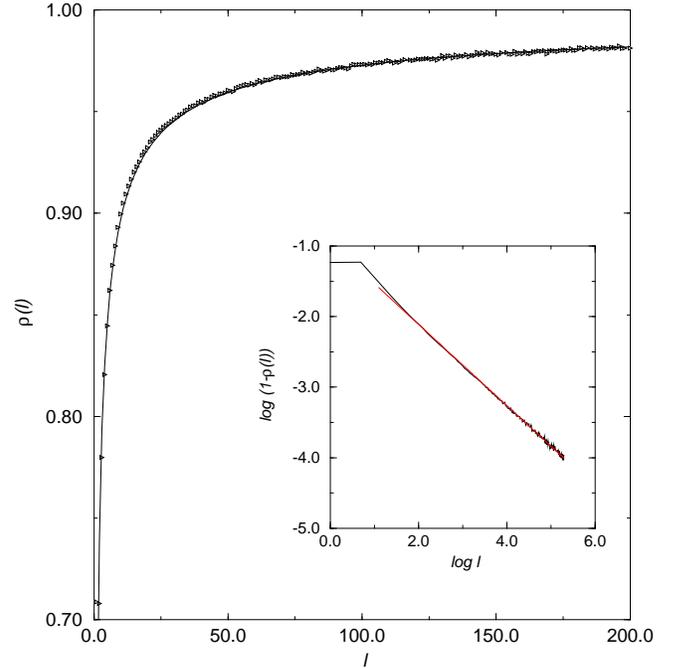}
\caption{\label{fig4}Density profile  $\rho (\ell )$ for $p=p^\star $
plotted against the distance $\ell $ from the top (triangles) and
the theoretical curve $\rho (\ell ) =1-0.39\ \ell ^{-0.578}$ (full
line). Inset: Data in double-logarithmic plot. The slope is $x =0.575 
\pm 0.005$. Within numerical error $x$ remains $p$-independent  
in the region $p \geq p^\star$. }
\end{figure}

\end{multicols}

\begin{references}
\bibitem[9]{correlations}The argument becomes exact with  the  rules 
modified such  that  heights at sites on the same layer are 
randomly redistributed fter each avalanche.
\end{references}
\end{document}